\documentstyle[prl,aps]{revtex}

\input epsf
\tighten

\begin{document}

\newcommand{\be}{\begin{equation}}
\newcommand{\ee}{\end{equation}}
\newcommand{\n}[1]{\label{#1}}
\newcommand{\ind}[1]{\mbox{\tiny{#1}}}
\newcommand{\ra}{\rangle}
\newcommand{\la}{\langle}
\newcommand{\para}{\parallel}
\newcommand{\ag}{\mbox{I \hspace{-0.82em} H}}
\def\bbox{{\,\lower0.9pt\vbox{\hrule \hbox{\vrule height 0.2 cm
\hskip 0.2 cm \vrule  height 0.2 cm}\hrule}\,}}
\newcommand{\half}{{1\over2}}
\newcommand{\beq}{\begin{equation}}
\newcommand{\eeq}{\end{equation}}
\newcommand{\bea}{\begin{eqnarray}}
\newcommand{\eea}{\end{eqnarray}}
\newcommand{\R}{\mbox{I \hspace{-0.82em} R}}
\newcommand{\req}[1]{Eq.(\ref{#1})}
\newcommand{\betabar}{\overline{\beta}}
\newcommand{\NB}{ {\bf  NB \quad}}

\title{Black Hole as a Point Radiator and  Recoil Effect  in the Brane World
}
\author{
Valeri Frolov and Dejan Stojkovi\'{c} }

\address{
Theoretical Physics Institute, Department of Physics, \ University of
Alberta,  Edmonton, Canada T6G 2J1
}

 \wideabs{
\maketitle
\begin{abstract}
 \maketitle
 \widetext
   A small black hole attached to a brane in a higher dimensional space
emitting quanta into the bulk may leave the brane as a result of
a recoil.   We construct a field theory model in which such a
black hole is described as a massive scalar particle with internal
degrees of freedom. In this model, the probability of transition between the different
 internal   levels followed by emission of massless
quanta is  identical to the probability of thermal emission calculated for the
Schwarzschild black hole. The discussed recoil effect implies
that the thermal emission of the black holes, which might be
created by interaction of high energy particles in colliders,
could be terminated and the energy non-conservation can be
observed in the brane experiments.
\end{abstract}                           }

\narrowtext

A common feature of all brane world models with large or infinite
extra dimensions \cite{ADD,RS} is that a lot of new interesting
phenomena can be expected  at the energy scale not much above the
energy currently available in accelerators. Probably the most
interesting and intriguing is the possibility of production of
mini black holes in future  collider and cosmic rays experiments.
Preliminary calculations \cite{Dim,Giddings} indicate that the
probability for creation of a mini black hole in near future
hadron colliders  such as LHC (Large Hadron Collider)  is so high
that they can  be called  ``black hole factories". Also, mini
black holes produced by ultra high energy cosmic rays could be
observed at the Auger Observatory before LHC starts operating
\cite{Feng}.

After the black hole is formed (either at LHC or the Auger
Observatory), it decays by emitting Hawking radiation. Thermal
Hawking radiation consists of  particles propagating along the
brane, and the bulk radiation. Usually the bulk radiation is
neglected since the total number of species which are confined on
the brane is quite large ($\sim 60$, see e.g. \cite{Dim,Myers}).
It should be noted that when the number of extra dimensions is
greater than 1 this argument may not work. Really, the number of
degrees of freedom of gravitons in the $(N+1)$-dimensional
space-time is ${\cal N}=(N+1)(N-2)/2$. For example, for $N+1= 7$
($3$ extra dimensions) ${\cal N}=14$.

It is also very important that mini black holes created in the
high energy scattering are expected to have high angular momentum
and to be extremal or close to them. For this reason, the effect
of superradiance must be important. The superradiant emission is
dominated by the particles with the highest spin, that is by
gravitons. This effect was studied in $(3+1)$-dimensional
space-time by  Don Page \cite{Page:76} (see also \cite{FrNo}). He
demonstrated that for an extremely rotating black hole the
probability of emission of a graviton by an extremely rotating
black hole is about 100 times higher than the probability of
emission of a photon or neutrino. We may expect that this
conclusion remains valid for higher-dimensional black hole so that
the bulk radiation may be comparable with (or even dominate) the
radiation along the brane.

But even for small number of extra dimensions the role of bulk
graviton emission might be important. As a result of the emission
of the graviton into the bulk space, the black hole recoil can
move the black hole out of the brane. Black hole radiation would
be terminated and an observer located on the brane would
register  the virtual energy non-conservation.

Since the problem in its complete scope is very complicated we
make some simplifying assumptions. We assume that the
characteristic size of extra dimensions is much larger than the
Schwarzschild radius of the black hole, so that we can
effectively describe the black hole by a higher dimensional
Schwarzschild solution.

In our model, the center-of-mass motion of black hole of mass $M$,
in $D=N+1$ space-time dimensions, is described by a scalar wave
function $\Phi$ with an action \be\n{2.1} W=-{1\over 2}\, \int\,
d^D x\, \left[ (\nabla \Phi)^2 +M^2\, \Phi^2\right]\, , \ee and
obeying the equation $ \Box \Phi -M^2 \Phi=0\, . $

We use the following mode decomposition for the quantum field
$\hat{\Phi}$ \be \hat{\Phi}(X^{A}) =\int\, {d^N {\bf P}\over
\sqrt{2\omega_{\bf P}}}\, {1\over (2\pi)^{N/2}}\, \left[
e^{-i\omega_{\bf P}t+i{\bf P\,X}}\, \hat{A}({\bf P})+  {\rm
h.c.}\right] \ee where h.c. stands for hermitian conjugate. The
bulk energy is $\omega_{\bf P}=\sqrt{{\bf P}^2+M^2}$.  Here and
later we use the following notation: $P^{A}$ ($A=1,...,N$) is
total bulk momentum which has components $p^i_{\para}$
($i=1,2,3$) along the brane, and $p^a_{\perp}$ ($a=1, ... ,N-3$)
in the bulk direction. We also denote $x^i$ coordinated along the
brane, and $y^a$ bulk coordinates. Thus, $ {\bf P\, X}={\bf
p}_{\para}\, {\bf x}+\bf{p}_{\perp}\, {\bf y}$.In the case where
there is no brane, all the space-like dimensions are equivalent.

Similarly, we write the mode decomposition for the bulk massless
scalar field $\varphi$

\be \n{2.7} \hat{\varphi}(X^{A})=\int\, {d^N {\bf K}\over
\sqrt{2\omega}}\, {1\over (2\pi)^{N/2}}\, \left[ e^{-i \omega
t+i{\bf K\,X}}\, \hat{a}({\bf K})+ {\rm h.c.} \right]\, . \ee
Here $\omega=K=|{\bf K}|$
is the bulk energy.

We choose the interaction action in the following form \be\n{2.8}
W_{\ind{int}}=\sum_{I\ne J}\, \lambda_{IJ}\, \int\, dX^D\,
\hat{\Phi}_I(X)\, \hat{\Phi}_J(X)\, \hat{\varphi}(X)\, . \ee

$\lambda_{IJ}$ is the coupling constant between the two different
internal states. We characterize these states $I$ by the value of
black hole mass $M_I$. Emission of quanta of a bulk field
$\varphi$ by the black hole changes its mass and hence provides a
transition $I\to J$ to the lower energy state $J$. The amplitude
of probability $A_{JK,I}$ of the particle (``black hole'')
transition from the initial state $I$ to the final state $J$ with
emission of a massless quantum $K$ is
 \begin{eqnarray}
A_{JK,I} &&=i\la {\bf P}_{J}, {\bf K}|W_{\ind{int}}|{\bf P}_I\ra=i\lambda_{IJ}
{2^{-3/2}\over (2\pi)^{N/2-1}}\, (\omega_{{\bf P}_I}\omega_{{\bf P}_J}
\omega)^{-1/2}\, \nonumber \\    &&      \times
\delta^{N}({\bf P}_I-{\bf P}_J-{\bf K})\,
\delta(\omega_{{\bf P}_I}-\omega_{{\bf P}_J}-
\omega)\, .     \end{eqnarray}

We assume that initially black hole is at rest, so that ${\bf
P}_I=0$ and $\omega_{{\bf P}_I}=M_I$. The probability for the
black hole to emit a quantum with energy $\omega$ per unit time
is
$$ p(\omega)\, ={(2\pi)^N\over \Delta t\, V_N}\,
\sum_{J}\, \int\, d^N\,{\bf P}_J \int d{\bf n}_K \,\omega^{N-1}\,
|A_{JK,I}|^2\, . $$
 Here $V_N$ is the space volume and $\Delta
t$ is the total time duration.  We also denoted ${\bf n}_K={\bf
K}/K$ so that $\int d{\bf n}_K$ is the averaging over direction
of emitted quanta ${\bf K}$.

After integration (the detailed calculations are given in
\cite{FS}), the final probability of emission of a massless
particle of energy $\omega$ per  unit time by our ``back hole''
of mass $M$ is \be\n{2.18} p(\omega|M)={\Omega_{N-1}\, \over 4\,
(2\pi)^{N-1}}{\omega^{N-2}\over M}\,\Lambda^2(M^2,M^2-2M
\omega)\, . \ee where $\Omega_{N-1}$ is a volume of a
($N-1$)-dimensional unit sphere and we omit the subscript ``I",
i.e. $M_I \equiv M$. We used $ M_I^2-M_J^2=\epsilon (I-J)$ and $
\lambda_{IJ}=\sqrt{\epsilon}\Lambda(M_I^2,M_J^2)$ in order to
provide the correct limit to the continuous mass spectrum case
($\epsilon \to 0$).

We demonstrate now that, for a special choice of the function
$\Lambda$, the probability rate (\ref{2.18}) coincides with the
probability of emission of scalar massless quanta of energy
$\omega$ by a black hole of mass $M$. For  $(N+1)$-dimensional
non-rotating Schwarzschild black hole, in the low frequency
approximation, we have (see e.g. \cite{GuKlTs:97,KaMa:02} and
references therein):

\be\n{2.37} P(\omega|M)={(\omega R_0)^{N-2}\over
2^{N-2}\Gamma^2(N/2)} {1 \over e^{\beta \omega R_0} -1 }\, . \ee
 where the Schwarzschild radius of $(N+1)$-dimensional
black hole $R_0 =\left[{16\pi G_* M \over
(N-1)\Omega_{N-1}}\right]^{1/(N-2)}$. $\beta = (T R_0)^{-1}= (4
\pi)/(N-2)$ is a dimensionless temperature and $G_* = 1/M_*^{N-1}
$ is a fundamental gravitational constant determined by a
fundamental energy scale $M_*$.

Comparing (\ref{2.18}) with (\ref{2.37}) one can conclude that if
we want a decaying massive particle $M$ to emit massless quanta
with the same probability as an evaporating black hole in the low
frequency approximation one must choose \be \label{lambda}
\Lambda^2(M^2,M^2-2M\omega)= {(2\pi)^{N-1}\over \pi^{N/2}\,
2^{N-3}\, \Gamma(N/2)}\,{ M\, R_0^{N-2}\over e^{\beta \omega
R_0}-1}\, , \ee

This result is valid for any ($N+1$)-dimensional black hole which
can be considered as a point radiator and is independent from the
brane world models. Since $R_0$ can be expressed in terms of $M$,
the right hand side of  eq. (\ref{lambda}) is a function of $M$
and $\omega$ which allows us to determine $\Lambda$.

 We note that the results concerning  the field theoretical model describing a
black hole as a point radiator, are quite general and are not
restricted to brane world models.

Suppose  now that there exists a brane  representing our physical
world embedded in a higher dimensional universe. For simplicity we
assume that the brane has a co-dimension 1, i.e.  there exist
only one extra dimension. The model can be easily generalized to
any number of dimensions. As earlier, we neglect the effects
connected with the spin and consider emission of the scalar
massless field. Moreover we use a simplified model to take into
account the effects connected with the interaction of the black
hole with the brane. Namely, we neglect the effects of the brane
gravitational field on the Hawking radiation, and use one
parameter, $\mu$, similar to a chemical potential, to describe the
interaction of the black hole with the brane.

The action is
 \be\n{3.1} W=-{1\over 2}\, \int\, d^D x\, \left[
(\nabla \Phi)^2 +U\, \Phi^2\right]\, , \ee
with $U=M^2-2\mu\, \delta(y)\, . $ The field $\Phi$  obeys the
equation $ \Box \Phi -U \Phi=0$ .

It is easy to see that there is only one level with positive
$\lambda=\mu^2$. A wave function of the corresponding bound state
is $ \Phi^{(0)}(y)=\sqrt{ \mu } \, e^{-\mu\, |y|} $.

For negative $\lambda$ the spectrum is continuous. We denote
$\lambda=-p_{\perp}^2$. Solving the scattering problem for the
$\delta$-like potential one obtains the following set of solutions

   \begin{eqnarray}
  \Phi^{(+)}_{p_{\perp}}(y)=
   e^{ i p_{\perp} y}-{\mu\over \mu +ip_{\perp}}\,e^{ -i p_{\perp} y} \ \ ,  y<0\ ,
    \nonumber  \\
{ip_{\perp}\over \mu +ip_{\perp}}\,e^{ i p_{\perp} y} \ \ , y>0 \
, \nonumber  \\
\Phi^{(-)}_{p_{\perp}}(y)= {ip_{\perp}\over \mu +ip_{\perp}}\,e^{
-i p_{\perp} y} \ \ ,
 y<0\ , \nonumber  \\
e^{ -i p_{\perp} y}-{\mu\over \mu +ip_{\perp}}\,e^{ i p_{\perp}
y} \ \ ,
 y>0 \ . \nonumber
\end{eqnarray}

The complete set of modes consists of two types of solutions. The
first type are solutions describing a black hole attached to the
brane which can freely propagate only along it: \be\n{3.12}
\Phi^{(0)}_{{\bf p}_{\para}}({X})={e^{-i\tilde{\omega}t}\over
\sqrt{2\tilde{\omega}}  \, (2\pi)^{(N-1)/2}}\, \Phi_0(y)\,
e^{i{\bf p}_{\para}{\bf x}_{\para}} \ee where $
\tilde{\omega}^2=\tilde{M}^2+p_{\para}^2 $ and
$\tilde{M}^2=M^2-\mu^2$. The second type of modes are bulk modes
of the form \be\n{3.14} \Phi^{(\pm)}_{{\bf P}}({X})={e^{-i\omega
t}\over \sqrt{2\omega}\, (2\pi)^{N/2}}\,
\Phi^{(\pm)}_{p_{\perp}}(y)\, e^{i{\bf p}_{\para}{\bf
x}_{\para}}    \ee where $\omega^2=M^2+p_{\para}^2+p_{\perp}^2$.

The field operator decomposition takes the form
\begin{eqnarray}
\hat{\Phi}(X) &&=\int\, d^{N-1}p_{\para}\, \left[
\Phi^{(0)}_{{\bf p}_{\para}}({X})\, \hat{B}({\bf p}_{\para})+
{\rm h.c.} \right] \nonumber \\ &&
 + \sum_{\pm}\, \int\, d^N {\bf P}\, \left[
\Phi^{(\pm)}_{{\bf P}}({X})\, \hat{A}_{\pm}({\bf P})+ {\rm h.c.}
  \right]\, .           \nonumber
\end{eqnarray}

For the scalar massless field we shall use the decomposition
(\ref{2.7}). This means that we neglect possible interaction of
scalar quanta with the brane.

The amplitude of probability $A_{JK,I}$ of the particle (``black
hole'') transition from the initial state $I$ to the final state
$J$ with emission of a massless quantum $K$ is $
A_{JK,I}^{off}=i\la {\bf P}_{J}, {\bf K}|W_{\ind{int}}|{\bf
P}_I\ra \, , $ where $W_{\ind{int}}$ is given by (\ref{2.8}). We
choose as the initial state $|I\ra$ a state of the black hole at
rest at the brane $ |I\ra =\hat{B}^{\dagger}({\bf
p}_{\para}=0)\,|0\ra $. We are interested in those final
states of a black hole when as a result of the recoil it leaves
the brane and moves in the bulk space, i.e. $ |J,K\ra
=\hat{A}^{\dagger}_{\pm}({\bf P})\,\hat{a}^{\dagger}({\bf
K})|0\ra \, . $

Thus, we have \begin{eqnarray} A_{JK,I\pm}^{off} && =i\lambda_{IJ}
{2^{-3/2}\over (2\pi)^{(N-1)/2}}\, (\tilde{M}_I \,\omega_{{\bf
P}_J} \omega)^{-1/2}\, \nonumber \\ && \delta^{N-1}({{\bf
p}_{\para}}_J+{\bf k}_{\para})\, \delta(\tilde{M}_I-\omega_{{\bf
P}_J}- \omega_k) C_{\pm}\nonumber \, , \end{eqnarray} where $
C_{\pm}= \int\, dy \, \Phi^{(0)}(y)\, e^{ik_{\perp}y}\,
\Phi^{(\pm)}_{p_{J\perp}}(y)$.

The total probability per unit time of the black hole emission
which results in the recoil taking the black hole away from the
brane is \be\n{3.23} w_{off}\, ={(2\pi)^{N-1}\over \Delta t
V_{N-1}}\sum_{J}\, \int\, d^N\,{\bf P}_J \int d{\bf K} \,
|A_{JK,I}^{off}|^2\, . \ee

Performing integrations using (\ref{lambda}) and the fact that the
main contribution to the integrals comes from the low frequency
modes (the high frequency ones are suppressed by a thermal
factor) give the following result for $w_{off}$ in $N=4$:
$$ w_{off}(\nu) ={\mu
\over 8 \pi^2 \nu } \int_0^{\infty}\, d\chi\,
{\chi^2\,(\sqrt{\chi^2+\nu^2}-\nu)\over \sqrt{\chi^2+\nu^2}}
{1\over \exp(\chi)-1 }\, . $$ in terms of dimensionless variables
$ \chi =a k$ and $ \nu=2 \mu a = 8 \sqrt{{2\pi \over 3}} \mu
\sqrt{G_* \tilde{M}}\, , $  where $a= 4 \sqrt{{2\pi \over 3}}
\sqrt{G_* \tilde{M}}$ and $k = \sqrt{k_{\para}^2+k_{\perp}^2}$ is
the absolute value of the total momentum of the emitted quantum
$K$. Also, $\tilde{M}_I=\tilde{M}$.

Calculation of the probability that the black hole remains  on
the brane is analogous to the previous calculation. The amplitude
of probability $A_{JK,I}$ for the particle (``black hole'') to
stay on the brane after emitting a massless quantum $K$ is $
A_{JK,I}^{on}=i\la {\bf P}_{J}, {\bf K}|W_{\ind{int}}|{\bf
P}_I\ra \, , $ where $W_{\ind{int}}$ is again  given by
(\ref{2.8}). We choose as the initial state $|I\ra$ a state of
the black hole at rest at the brane, $ |I\ra
=\hat{B}^{\dagger}({\bf p}_{\para}=0)\,|0\ra \, . $ We are
interested in those final states of a black hole when as a result
of the recoil it remains on the brane, i.e. $ |J,K\ra =
\hat{B}^{\dagger}({\bf p}_{\para} \neq 0)
\,\hat{a}^{\dagger}({\bf K})|0\ra \, . $

Repeating  calculations we obtain (for $N=4$):

 \be\n{4.21} w_{on}(\nu)= {8 \mu \over \pi} \int_0^{\infty}\,
d\chi\, {\chi^2\, \over \sqrt{\chi^2+\nu^2}} {1\over \exp(\chi)-1
} \ee
 in the same dimensionless variables as for $w_{off}$.

\begin{figure}
\centerline{\epsfxsize = 0.70 \hsize \epsfbox{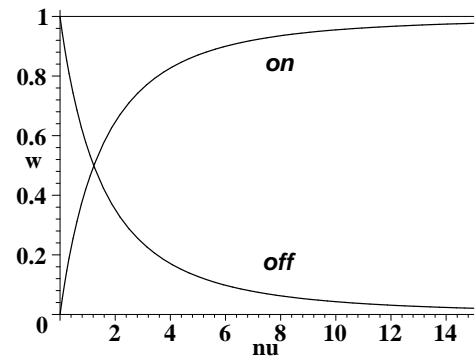}}
\caption{Functions $w_{off}(\nu)/w$ and $w_{on}(\nu)/w$.}
\label{w}
\end{figure}

It is easy to see that the sum $ w=w_{off}(\nu)+w_{on}(\nu) $
does not depend on $\nu$. In fact $w$ coincides with the total
probability of  emission of  a massless field quantum by the
5-dimensional black hole. \be\n{4.23} w=\int_0^{\infty}\, d\omega
\, P(\omega|M)\, , \ee where $P(\omega|M)$ is given by
(\ref{2.37}). The plots of the functions $w_{off}(\nu)/w$ and
$w_{on}(\nu)/w$ are shown at Fig.~\ref{w}. Analysis shows that,
for large $\nu$ (or $\mu$), the probability $w_{off}$ falls off as
$1/\nu$, while for small $\nu$ (or $\mu$) the (normalized)
probability $w_{off}$ goes to $1$.

In literature, the process of evaporation of mini black holes in
brane world models is well studied. The pattern of such a black
hole decay is markedly different from any other standard model
event. There exist very detailed calculations estimating the
energy spectrum and ratio between emitted particles (leptons,
fotons, hadrons etc.)\cite{Giddings} . It is also claimed that,
because of very little missing energy, the determination of the
mass and the temperature of the black hole may lead to a test of
Hawking's radiation. The recoil effect discussed above may change
some of these predictions.  Certainly the most important
observable effect of a black hole recoil is a suddenly disrupted
evaporation and local non-conservation of energy.

We developed a phenomenological model for description of the
recoil effect. This model contains an important parameter $\mu$
which play the role of the chemical potential. In the general
case, $\mu$ depends on the mass of a black hole $M$ and the
tension $\sigma$ of the brane, $\mu=\mu(M,\sigma)$.
Unfortunately, to determine this dependence is not an easy task.
One can expect that in the limit $\sigma\to 0$, that is for a
test brane, $\mu\to 0$. In this case, it is very likely that the
black hole leaves the brane as soon as emits first quanta with
non-zero bulk momentum (see Fig. \ref{w}).

We can also estimate $\mu$ for small $\sigma$ as follows.
Consider two different states: first, a brane with a black hole of
radius $R_0$, and the second, when the same black hole is out of
the brane. The second configuration has extra energy $\Delta E
\sim \sigma R_0^3$ (for 1 extra dimension). One can identify
$\Delta E$ with $M-\tilde{M}$. This gives $\mu^2\approx 2M\Delta
E\, , $ (see eq. (\ref{3.12})) or $ \mu\sim \sigma^{1/2}\,
R_0^{5/2}\sim \sigma^{1/2}\, M_0^{5/4}\, . $

In the other limit of infinitely heavy brane, or a brane with
$Z_2$ symmetry, the process of a black hole leaving the brane
reminds to a black hole splitting into two symmetric black holes
in the ``mirror'' space. Classically this process is forbidden in
a higher dimensional space-time for the same reason as in
$(3+1)$-dimensional space-time in connection with non-decreasing
property of the entropy. In the presence of cosmological
constant, such an effect may become possible as a tunneling
process. These arguments show that in this case $\mu \rightarrow
\infty$ or is exponentially large, and the recoil effect is
suppressed. This feature could help in distinguishing between the
two different scenarios of compact and infinite extra dimensions.
Also, if the general conclusion that the recoil effect may be
important for branes of small tension is correct, it opens an
interesting possibility of using the experiments with decay of
mini black holes to put restrictions on the brane tension.

The above order of magnitude estimates of the parameter $\mu$ do
not follow from the exact calculations and therefore  further
investigation is required\footnote{Our attention was drown by Don
Page who pointed out that more detailed analysis shows that the
estimate for $\mu$ given above is quite valid in the case of the
brane of a co-dimension $3$ and higher (three or more extra
spatial dimensions). For co-dimensions $1$ and $2$ result might
depend on a regime in which we extract the black hole, i.e.
whether it is an adiabatic process or not.}.

\bigskip

     \vspace{12pt} {\bf Acknowledgments}:\ \  The authors are grateful
to Don Page and Glenn Starkman for stimulating discussions. This work was partly
supported  by  the Natural Sciences and Engineering Research
Council of Canada. The authors are grateful to the Killam Trust
for its financial support.


\begin{thebibliography}{9}







\bibitem{ADD}
N. Arkani-Hamed, S. Dimopoulos and G. Dvali, Phys. Lett. {\bf B429},
263 (1998); I. Antoniadis, N. Arkani-Hamed,  S. Dimopoulos and G. Dvali,
Phys. Lett. {\bf B436}, 257 (1998).






\bibitem{RS}
L. Randall and R. Sundrum, Phys. Rev. Lett. {\bf 83}, 3370
(1999); {\it ibid} 4690 (1999);



\bibitem{Dim} S. Dimopoulos, G. Landsberg, Phys. Rev. Lett. {\bf 87} 161602 (2001)

\bibitem{Giddings}
 D. M. Eardley and S. B. Giddings, {\it gr-qc/0201034} ; S. B. Giddings and S. Thomas, Phys. Rev. {\bf D65} 056010 (2002)



\bibitem{Feng} J.  L. Feng, A. D. Shapere,  Phys. Rev. Lett. {\bf 88} 021303 (2002)


\bibitem{Myers} R. Emparan, G. Horowitz, R. C. Myers, Phys. Rev. Lett.
{\bf 85} 499 (2000)









\bibitem{Page:76} D. N. Page, Phys. Rev. {\bf D13} 198 (1976);
    Phys. Rev. {\bf D14} 3260 (1976).

\bibitem{FrNo} V. Frolov and I. Novikov. {\em Black Hole Physics: Basic
Concepts and New Developments} (Kluwer Academic Publ.), 1998.




\bibitem{FS}   V. Frolov and D. Stojkovic, hep-th/0206046

\bibitem{GuKlTs:97} S. S. Gubser, I. R. Klebanov and A. A. Tseytlin,
Nucl. Phys. {\bf B 499}, 217 (1997).

\bibitem{KaMa:02} P. Kanti and J. March-Russel, {\it hep-th/0203223}




\end{thebibliography}
\end{document}